\documentclass[twoside,british,a4paper,twocolumn]{article}

\usepackage{amsmath}
\usepackage{amsfonts}
\usepackage{amssymb,upref}
\usepackage{tgheros} 
\usepackage{tgtermes} 
\usepackage{anyfontsize}
\usepackage{xcolor}
\usepackage{graphicx}
\usepackage{enumitem}
\usepackage{subfig}
\usepackage{siunitx}
\usepackage{ulem}

\usepackage{nccmath,amsmath}
\usepackage{amsfonts}
\usepackage{mathtools}
\usepackage{tensor}
\usepackage{scalerel}

\usepackage{siunitx}
\usepackage{amsmath}
\usepackage{physics} 
\usepackage{bm}

\usepackage{balance}

\usepackage[utf8]{inputenc}
\usepackage[T1]{fontenc}

\usepackage{isodate}
\cleanlookdateon

\usepackage{lettrine}

\pdfmapfile{=montserrat.map} 

\usepackage{lipsum}

\usepackage{newtxmath}

\usepackage{geometry}
\usepackage{fancyhdr}

\fancypagestyle{firstpagestyle}
\usepackage{titlesec}
\titleformat{\section}[block]{\bfseries\upshape\sffamily\boldmath}{}{0.em}{}
\titlespacing*{\section}{0pt}{0.8em plus 0ex minus 0ex}{0em plus 0.ex}

\usepackage{titling}
\pretitle{
\begin{minipage}{\textwidth}%
\raggedright\huge\bfseries\upshape\sffamily\boldmath
\color[rgb]{0,0,0.8}%
}
\posttitle{%
\end{minipage}
\vskip 0.6em%
}
\preauthor{\noindent 
\begin{minipage}
{\textwidth} \large\mdseries\sffamily}
\postauthor{
  \end{minipage} 
  \vskip 0.4em
  {
   \sffamily
   \color{black!80}
   \noindent \small
   \address
   \par 
   \authoremail
   }
   \par \vskip 0.4em
  }
\predate{ \noindent \hspace{-0.4em} \sffamily\small}
\postdate{\vskip 0.0em}

\usepackage{caption}
\DeclareCaptionFont{cfs}{\fontsize{8.5}{10.25}\selectfont}
\DeclareCaptionLabelSeparator{vline}{\;|\;}
\usepackage{tcolorbox}
\definecolor{abstractboxcolor}{cmyk}{0.1,0,0,0}
\newtcolorbox{abstractbox}{
  arc=0pt,
  boxrule=0pt,
  colback=abstractboxcolor,
  boxsep=0.5em,
  left=0pt, right=0pt, bottom=0pt, top=0pt,
  width=\columnwidth
}
\makeatletter
 \def\@textbottom{\vskip \z@ \@plus 1pt}
 \let\@texttop\relax
\makeatother
\renewenvironment{abstract}{
   \noindent
   \begin{minipage}{\textwidth}
   \upshape\sffamily \bfseries
   \fontsize{9}{11.5}\selectfont
  }{
   \end{minipage} 
   \vskip 2.0em
  }

\usepackage[numbers,sort&compress]{natbib}

\makeatletter \def\NAT@def@citea{\def\@citea{\NAT@separator\,}} \makeatother 
\usepackage{etoolbox}
\apptocmd{\sloppy}{\hbadness 10000\relax}{}{}

\usepackage[%
  bookmarks=true,
  colorlinks,
  linkcolor=blue,
  urlcolor=blue,
  citecolor=blue,
  plainpages=false,
  pdfpagelabels,
  final,
  breaklinks=true
]{hyperref}
\hypersetup{
pdftitle={Nonadiabatic Floquet–Bloch coupling in high-harmonic generation from magnesium oxide}, 
pdfauthor={Cong Zhao et al.},
pdfkeywords={%
Floquet-Bloch states, 
high-harmonic generation, 
nonadiabatic coupling}
}

\title{
Floquet-engineering unveiled by high-harmonic generation
}

\newcommand\shorttitle{Floquet-engineering unveiled by high-harmonic generation}

\author{%
Cong Zhao\textsuperscript{1$\,*$}, 
Lucie Jurkovičová\textsuperscript{2,3}, 
Xiaozhou Zou\textsuperscript{1}, 
Benjamin T. Q. Miller\textsuperscript{1},
Suzan Canbas\textsuperscript{1},
Zakaria Dahbi\textsuperscript{1},
Martin Albrecht\textsuperscript{2,3},
Ondřej Finke\textsuperscript{2,3},
Jaroslav Nejdl\textsuperscript{2,3},
Margarita Khokhlova\textsuperscript{1},
Ondřej Hort\textsuperscript{2},
Fabrice Catoire\textsuperscript{4$\,*$}, and
Amelle Zaïr\textsuperscript{1$\,*$}
}
\newcommand\shortauthor{Cong Zhao et al.}
\newcommand\address{%
\textsuperscript{1}Attosecond Quantum Physics Laboratory, King's College London, Strand, London, WC2R 2LS, United Kingdom \\
\textsuperscript{2}ELI Beamlines Facility, the Extreme Light Infrastructure ERIC, Za Radnicí 835, 252 41 Dolní Břežany, Czechia \\
\textsuperscript{3}Czech Technical University in Prague, FNSPE, Břehová 7, 115 19 Prague 1, Czechia \\
\textsuperscript{4}CELIA, CNRS-CEA-Université de Bordeaux, 351 Cours de la Libération, Talence, F-33405, France \\
}
\newcommand\authoremail{$^*$cong.zhao@kcl.ac.uk; fabrice.catoire@u-bordeaux.fr; amelle.zair@kcl.ac.uk}

\graphicspath{{./}{./figure/}{../figure/}}

\usepackage{physics}

\usepackage{textcomp}

\begin{document}

\twocolumn[
\begin{@twocolumnfalse}

\maketitle
\thispagestyle{firstpagestyle}

\begin{abstract}

Ultrafast optical control of solids has uncovered new phenomena and advanced non-equilibrium condensed-matter physics, where photon-dressed electronic states---Floquet–Bloch states (FBSs)---emerge under a strong oscillating laser field, also known as Floquet engineering. Although FBSs have been extensively investigated using time- and angle-resolved photoemission spectroscopy, direct evidence of their role in high-harmonic generation spectroscopy (HHGS) has remained elusive. Here, we present combined experimental and theoretical evidence that FBSs can be probed by HHG emission in the wide-bandgap solid magnesium oxide (MgO) driven by few-cycle near-infrared pulses. Experimentally, we observe clear evidence of FBSs in the HHG yield dependence on the crystal orientation. This specific feature is attributed to nonadiabatic coupling between FBSs and conduction bands near the Brillouin-zone edge, where the strong laser field transiently breaks time-reversal symmetry. We have confronted the experimental findings with numerical solutions of the time-dependent Schrödinger equation, which reproduce the new feature and confirm its Floquet origin. The theoretical results show a coupling inducing a local band-structure renormalization and Floquet-like hybridization under strong-field excitation. It also shows that FBS nonadiabatic dynamics persist in the strong-field regime, establishing HHGS as a powerful probe of ultrafast light-induced band hybridization in solids.
\end{abstract}

\vspace{-2mm}

\end{@twocolumnfalse}
]

\lettrine[lines=3,lhang=0.15,nindent=0.1em]{U}{ltrafast} optical manipulation of solids pushes condensed-matter physics into new regimes, uncovering emergent phenomena that remain unreachable at equilibrium~\cite{hu2014optically,matsunaga2014light,stojchevska2014ultrafast,mitra2024light}. A prominent example is the emergence of Floquet–Bloch states (FBSs)~\cite{faisal1997floquet}, formed when a time-periodic light field drives a solid. Complex dynamics through virtual photon absorption and emission processes lead to a modification of the electronic eigenstates, giving rise to FBSs.
In the quasienergy spectrum, their spectroscopic manifestation as replica bands shifted by integer multiples of the photon energy, known as Floquet–Bloch sidebands (FBSBs)~\cite{Oka2025,wang2013observation}, which retain the orbital character of the original Bloch bands.

By now, considerable efforts have been devoted to FBSBs in the emerging field of Floquet engineering~\cite{rudner2020band,zhang2022angle,oka2019floquet,dai2022poincare}, particularly in two-dimensional materials with topological hexagonal structures and narrow band gaps. In these systems, the characteristic Dirac-cone energy–momentum dispersion can be periodically driven by a mid-infrared pump pulse, generating FBSBs that intersect the Dirac cone. This interaction leads to the nonadiabatic hybridization or coupling of the band structure~\cite{rudner2020band}, resulting in the opening of light-induced dynamic gaps at the Dirac point that the driving field can control. Such phenomena have been successfully observed using time- and angle-resolved photoemission spectroscopy (Tr-ARPES) in various materials, such as monolayer graphene~\cite{choi2025observation,merboldt2025observation,Oka2025}, black phosphorus~\cite{Zhou2023, Bao2024, Zhou2025, Fan2025}, cadmium selenide nanoplatelet~\cite{li2024observation}, Weyl semimetal~\cite{sirica2022photocurrent}, topological insulator~\cite{mahmood2016selective,wang2013observation,ito2023build,Bielinski2025}, and transition metal dichalcogenide~\cite{aeschlimann2021survival,Fragkos2025}. Through the prism of Floquet engineering, these studies reveal how non-equilibrium laser dressing can break time-reversal symmetry (TRS), giving rise to anomalous quantum transport phenomena such as the light-induced Hall effect~\cite{oka2009photovoltaic,mciver2020light,Oka2025}.

In principle, the contribution of FBSs should manifest across different types of photon emission spectra in solids—for example, in high-harmonic generation spectroscopy (HHGS), which is typically governed by intraband current and interband polarization mechanisms~\cite{golde2008high,vampa2015all,vampa2017merge, Catoire2018}. In the past few decades, HHGS~\cite{cavaletto2025attoscience,heide2024ultrafast,goulielmakis2022high,ghimire2019high,tao2016direct,ghimire2011observation,schubert2014sub} has been served as a promising tool for probing the optical band dispersion~\cite{vampa2015all,luu2016high,lanin2019high}, the crystal's dynamical symmetry~\cite{yue2022signatures,neufeld2019floquet,you2017laser}, the Berry phase character~\cite{luu2018measurement,uzan2024observation}, and
the electron correlations in materials~\cite{silva2018high,uchida2022high,freudenstein2022attosecond,chang2024many}. 
However, direct observation of FBSBs in HHGS remains a challenge. One reason is that the momentum-integrated nature of HHG averages out Floquet features and can suppress them through channel-closing effects~\cite{jin2019contribution}. Furthermore, the reliance on a single-color driving field limits the ability to spectrally disentangle Floquet-induced contributions through each wavelength. Nevertheless, HHGS still provides a powerful platform for probing FBSs. 
Although FBSBs may not be directly observable, their fingerprints are expected to manifest 
in the strong-field response and play a decisive role in shaping the structure and dynamics of high-order harmonics in solids. For example, high-order sideband generation in MoS$_2$ has been attributed to Raman-type scattering between Floquet states driven by mid-infrared light~\cite{nagai2020dynamical}. Floquet states have also been shown to govern attosecond optical responses in dielectrics through nonadiabatic virtual interband coupling in the light-dressed solid~\cite{dolso2025attosecond}. FBSs have further been shown to strongly modify optical nonlinearities through light-induced TRS breaking~\cite{ikeda2018floquet,jin2019contribution,mitra2024light}, leading to giant modulation in low-order harmonic generation~\cite{shan2021giant} and pronounced spectral shifts~\cite{dimitrovski2017floquet}. On the other hand, unlike Tr-ARPES, which typically requires low temperatures and weak excitation in few-layer materials to maintain near-equilibrium conditions, solid-state HHGS operates in the strong-field regime and enables access to strongly-driven non-equilibrium dynamics in bulk crystals.

While FBSs have been increasingly recognized in HHG under strong-field conditions, their manifestation via nonadiabatic Floquet engineering remains largely unexplored. Establishing experimental evidence for Floquet-mediated nonadiabatic coupling would represent an important step towards the direct observation of FBSBs in HHGS, particularly when the crystal azimuthal angle scan ($\alpha$ scan) is employed to unveil band-dependent dynamics.

In this work, we uncover the role of FBS–mediated nonadiabatic coupling in shaping HHGS in solids. This dynamic is addressed in the wide-bandgap dielectric --- magnesium oxide (MgO) --- and driven by an intense few-cycle near-infrared (NIR) field.
The Floquet parameter ($\mathcal{F}$) in the experiment exceeds unity~\cite{dunlap1986dynamic}, indicating that transient light-dressed FBSBs can emerge on this timescale. Although the standard measured HHG spectrum hampers direct FBSB observation, using $\alpha$ scan, we identify an additional spectral feature -- within the energy range of 18-21 eV (below harmonic order 13, HH13) -- attributed to FBSB due to nonadiabatic coupling between FBSs and the conduction bands of MgO, analogous to Floquet engineering.
The coupled bands open a dynamical energy gap and are separated by one laser photon energy, resulting in a transition photon emission located around HH12 (an FBSB). This signal follows the morphology of MgO’s band gap structure as the crystal azimuth angle $\alpha$ is rotated, giving rise to a distinctive arc-like feature in the HHG spectrum. Such a nonadiabatic process is predominantly associated with crystal momenta near the Brillouin-zone (BZ) edge, where multiple conduction bands converge and interband polarizations become dominant. Previous studies using infrared pump–probe techniques have demonstrated the Landau–Zener–type transitions in this momentum range~\cite{uzan2020attosecond,uzan2022observation}. Here we employ a strong few-cycle NIR pulse to directly probe the strong-field-induced nonadiabatic coupling between FBSs and electronic bands, showing that HHGS captures dynamical Floquet signatures in the strong-field regime and that FBS-driven nonadiabatic coupling is a key mechanism of interband HHG in solids.

We begin by reporting an experimental investigation of Floquet-state dynamics in HHGS of MgO, with particular emphasis on the BZ edge region. By combining HHG measurements with time-dependent Schrödinger equation (TDSE) simulations and Floquet-dressed band-structure models derived from density functional theory (DFT), we identify distinct spectroscopic signatures of non-equilibrium electron dynamics driven by strong light–matter interaction.
\section*{Observation of Floquet signatures in HHG from MgO}
\begin{figure*}[ht!]
\centering\includegraphics[width=1\textwidth]{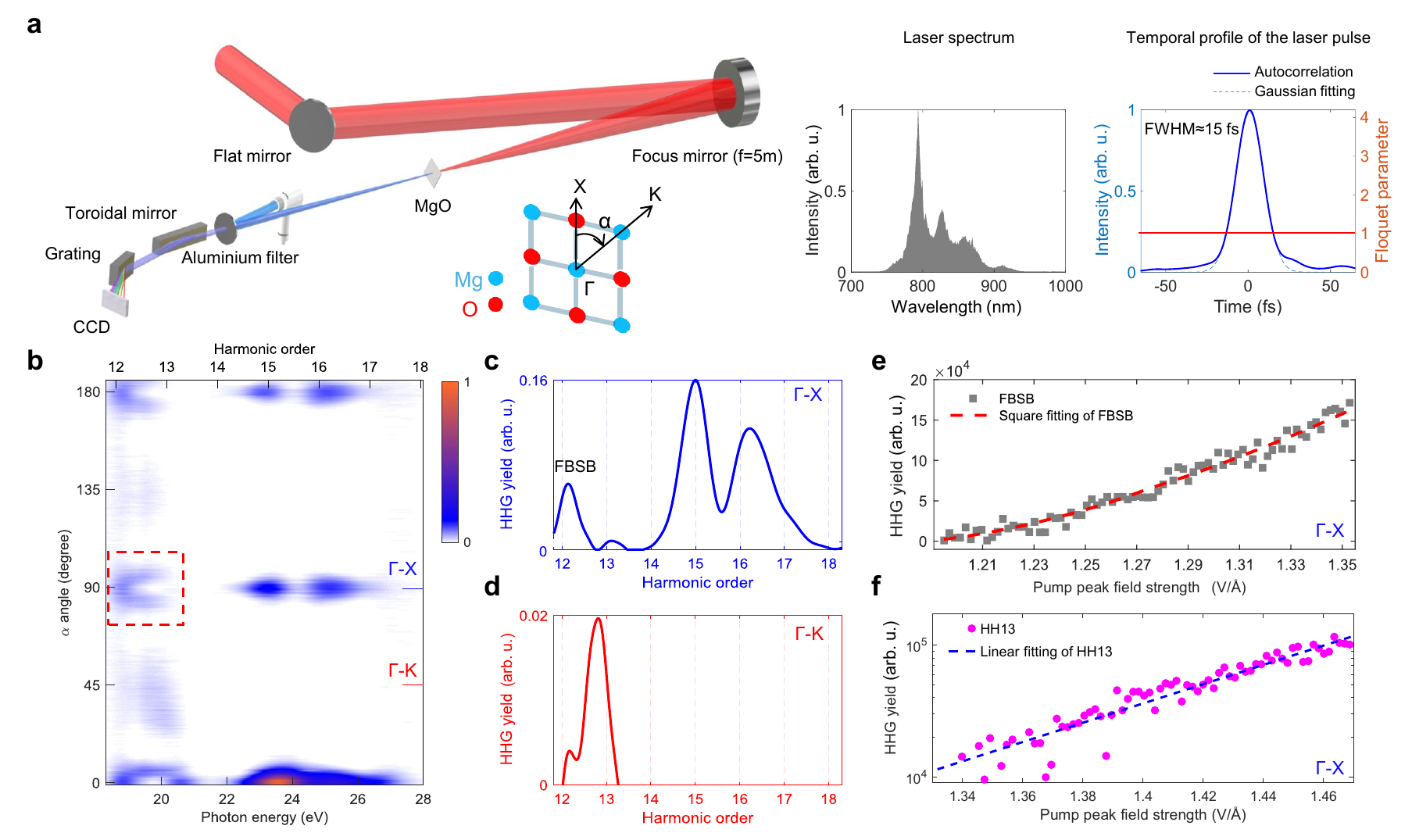}
\caption{Experimental results for HHG in MgO. \textbf{a}~Experimental setup for HHGS measurement. A NIR laser field is focused onto the (100) surface of MgO. The inset illustrates the crystal structure of MgO, where the $\Gamma$–X axis lies along the Mg–O bond direction, and the $\Gamma$–K axis is rotated by 45$^\circ$ relative to it. The azimuthal angle $\alpha$ is defined as 0$^\circ$ when the laser polarization aligns with the $\Gamma$-X direction. The right panel shows the laser spectrum and temporal profile: the pulse is centered at 800 nm with a FWHM duration of 15 fs. The red line marks the condition where the Floquet parameter is equal to 1; excitation above this threshold indicates the formation of FBSs. \textbf{b}~HHGS orientation scan, defined by the azimuthal angle $\alpha$, of the HHG spectrum normalized to the maximum intensity of HH15 at $\alpha=\SI{0}{\degree}$; dashed rectangular shape defines the area of interest around \SI{18}-\SI{20}{eV} with the arc-like distribution of FBSB around the $\mathrm{\Gamma}$-X direction. \textbf{c,d}~HHG spectra at (\textbf{c})~$\alpha=\SI{90}{\degree}$~($\mathrm{\Gamma}$-X) showing the presence of multiple harmonic orders (HH13, HH15, HH17), and (\textbf{d})~$\alpha=\SI{45}{\degree}$~($\mathrm{\Gamma}$-K) showing single odd HH13, while the FBSB around HH12 appears at both angles, induced by FBS. \textbf{e}~Total yield of the FBSB, integrated over the full spectral linewidth, as a function of the pump peak field strength along the $\mathrm{\Gamma}$–X direction. The red dashed line shows a square fit, consistent with the expected scaling of a first-order FBSB in the weak-field regime. \textbf{f}~Total yield of HH13, integrated over the full harmonic linewidth, as a function of the pump peak field strength along the $\mathrm{\Gamma}$–X direction. The blue dashed line represents a linear fit, indicating a distinct scaling behavior of the interband harmonics.}
\label{fig2}
\end{figure*}

We first present the experimental observation of a distinct spectral feature in the HHG response of MgO, which serves as the central result of this work. In the experiment, this feature appears as a distinct, field-controllable fingerprint in the HHG spectrum.
To probe this feature in the strong-field regime, we drive a 100~$\mu$m-thick MgO crystal with the (100) surface using an intense near-infrared pulse and record the HHG spectra in transmission geometry.

The experimental configuration is shown in Fig.~\ref{fig2}a (see Methods for details), and the pulse temporal profile is plotted in the right panel. The driving field is centered at $800\,\mathrm{nm}$ ($1.55\,\mathrm{eV}$), with a peak intensity of $3\times10^{13}\,\mathrm{W/cm^2}$ and a duration of $15\,\mathrm{fs}$ (6 optical cycles FWHM, 14 cycles total). Under these conditions, $\mathcal{F}$ reaches 4.2, as determined from Eq. (2), placing the system in a strongly nonadiabatic regime with substantial FBS participation. For the temporal profile measured by the autocorrelation, the red solid line marks the intensity at which the Floquet parameter equals unity, above which light-dressed FBSs can form. The inset illustrates the crystal structure of MgO, where the $\Gamma$–X axis lies along the Mg–O bond direction and the $\Gamma$–K axis is rotated by $45^\circ$ relative to it. The azimuthal angle is defined as $0^\circ$ when the laser polarization is aligned with the $\Gamma$–X direction.

The measured HHG spectrum as a function of the azimuthal angle $\alpha$ is shown in Fig.~\ref{fig2}b. We focus on the energy range around the second plateau of the spectrum~\cite {you2017laser, uzan2022observation, allegre2025extension} in the energy region between \SI{18}{eV} and \SI{28}{eV}, where multiple CBs contribute significantly to HHG near the BZ edge. Besides, at an initial azimuthal angle of $\SI{0}{\degree}$, we observe a slight difference in the HHG signal, which appears stronger and where the arc-like structure appears more circular than at $\SI{90}{\degree}$ and $\SI{180}{\degree}$. We attribute this difference mainly to extrinsic experimental factors rather than an intrinsic MgO response. The MgO crystal cutting-angle tolerance is $~\pm\SI{2}{\degree}$, which can average the effective crystallographic orientation. In addition, laser-induced cumulative heating during the alpha scan, can modify HHG signal strength.

In Fig.~\ref{fig2}b, the $\alpha$ angle scan of the HHG spectra exhibits generally four-fold symmetry, where the HHG is generated primarily at azimuthal angles $\alpha=\SI{90}{\degree}$~($\mathrm{\Gamma}$-X) and $\alpha=\SI{45}{\degree}$~($\mathrm{\Gamma}$-K) with a period of $\SI{90}{\degree}$. This symmetry-dependent response arises from electron recombination, which depends on the crystal orientation~\cite{you2017anisotropic} and imprints the crystal structure. Along the $\mathrm{\Gamma}$-X (Mg-O) direction (Fig.~\ref{fig2}c), odd harmonics (HH13, HH15, and HH17) are observed, with an additional photon emission located around HH12 attributed to an FBSB, which is nominally forbidden by inversion symmetry. In contrast, along the $\mathrm{\Gamma}$-K (Mg-Mg) direction, only HH13 persists, while an additional weak feature near HH12 is also observed.

The four-fold rotational symmetry is preserved for most harmonics, but breaks down with the FBSB, where an anomalous arc-like feature emerges that is absent in other HH orders. This feature, appearing prominently along the $\Gamma$–X direction lies at an energy position around HH12 that is nominally forbidden by symmetry. When the crystal rotates, the feature undergoes a continuous blue shift and forms an arc-shaped trace (red dashed box in Fig.~\ref{fig2}b), identifying it as a transition signal rather than a conventional high-harmonic emission. In order to further confirm the origin of this feature, we analyze the dependence of the total yield on the pump peak field strength in the $\mathrm{\Gamma}$-X direction for the FBSB, compared with the spectrally closest odd HH13. For HH13 (see Fig.~\ref{fig2}f), the yield scales linearly with the peak field strength, indicating a perturbative dependence on the driving field, with the linear regime emerging from a threshold of $1.33\,\mathrm{V/\mathring{A}}$. The blue dashed line shows the linear fit to the measured HH13 signal. In contrast, the signal around HH12 (grey squares in Fig.~\ref{fig2}e) exhibits a pronounced increase in yield with the pump peak field strength. The fitted trend (red dashed line) follows a square scaling. Within Floquet--Bloch theory~\cite{wang2013observation,saathoff2008laser,miaja2009ultrafast}, the intensity of the $n$-th sideband in a photon emission spectrum reads
\begin{equation}
I_{n}(F_0) \propto \left| J_{n}\!\left( \frac{e a F_0}{\hbar \omega} \right) \right|^{2},
\end{equation}
where $J_{n}$ is the $n$-th order Bessel function. In the weak-field regime, the first-order Bessel function follows the asymptotic form: $J_{1}(x) \approx x/2$. This leads to $I_{1}(F_0) \propto F_0^{2}$, in agreement with the measured square dependence. The strong correspondence between the yield around HH12 and the square-law fit thus supports its assignment as a first-order FBSB. We exclude the signal around HH16 from further analysis, as it originates from transitions involving a higher conduction band (Fig.~\ref{fig31}a), which lies outside the scope of this work. We attribute this feature to an FBSB mediated by higher conduction bands, and therefore do not consider it in the discussion below.

\begin{figure*}[ht!]
\centering\includegraphics[width=1\textwidth]{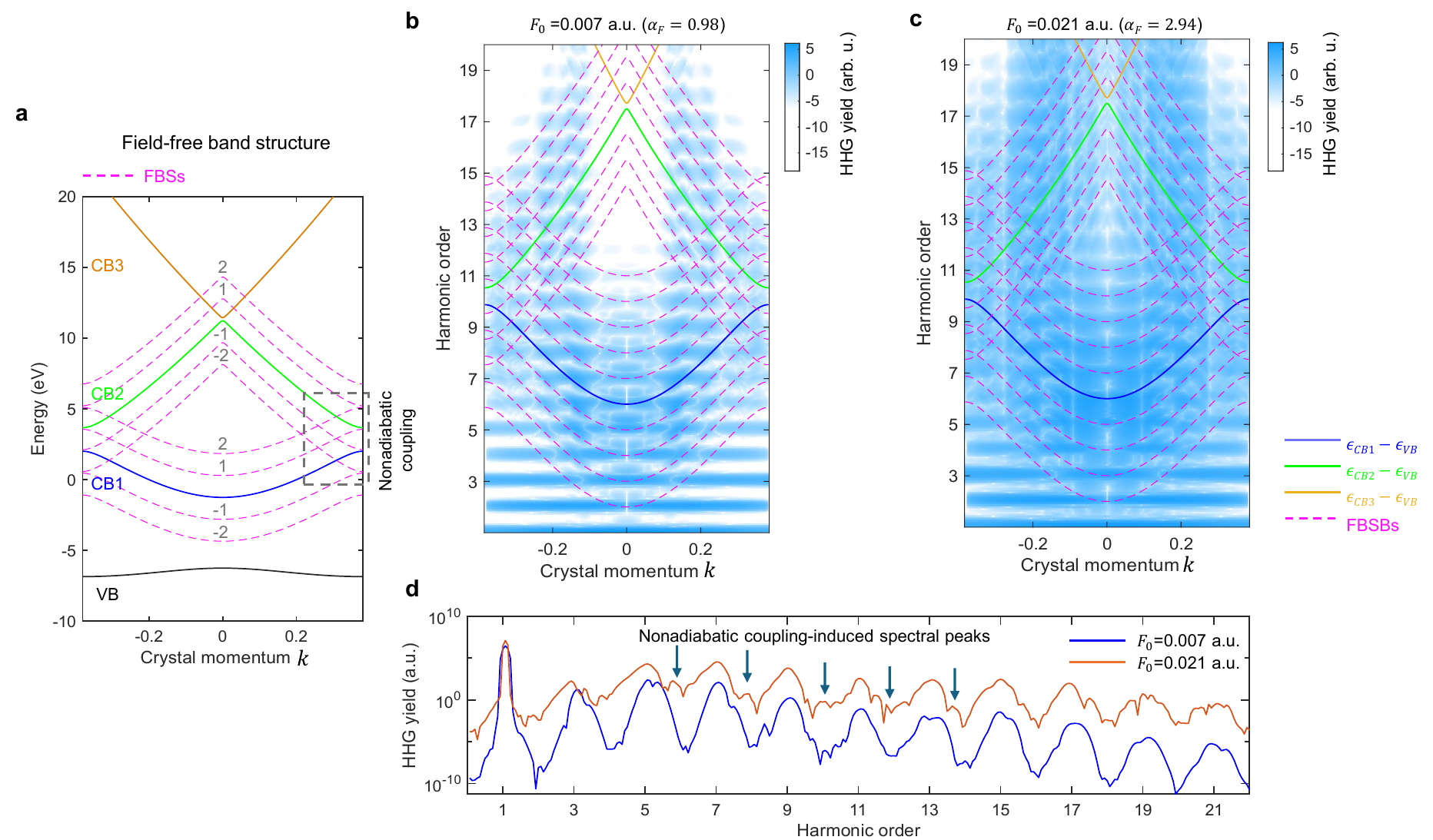}
\caption{\textbf{a}~Field-free band structure of the model potential used in the TDSE simulations, showing the valence band (VB, black curve) and three conduction bands (CB1, blue curve; CB2, green curve; CB3, yellow curve). The FBSs of CB1 and CB2 are plotted as purple dashed curves. 
\textbf{b}~Calculated HHG spectrum as a function of crystal momentum $k$ for a peak field strength of $F_0 = 0.07$ a.u., corresponding to a Floquet parameter $\mathcal{F} \approx 1$ (0.98). The formation of FBSBs (purple dashed curves) is clearly observed. Three interband transition trajectories are identified, corresponding to three momentum-resolved band gaps: gap 1 ($\epsilon_{\mathrm{CB1}} - \epsilon_{\mathrm{VB}}$, blue curve), gap 2 ($\epsilon_{\mathrm{CB2}} - \epsilon_{\mathrm{VB}}$, green curve), and gap 3 ($\epsilon_{\mathrm{CB3}} - \epsilon_{\mathrm{VB}}$, yellow curve). The FBSBs trace the band-gap dispersions of gap 1 and gap 2, with multiple photon-replica branches. These FBSBs intersect with successive harmonic orders, revealing nonadiabatic coupling between the FBSs and conduction bands, particularly at the BZ edge, which indicates the emergence of light-dressed hybridization between electronic states.  
\textbf{c}~HHG spectrum for a larger peak field strength of $F_0 = 0.021$ a.u., corresponding to a Floquet parameter $\mathcal{F} = 2.94$. In this case, the nonadiabatic coupling between the FBSs and electronic bands is more pronounced, and the overlap of FBSBs with the three band gaps becomes intrinsic. 
\textbf{d}~HHG spectrum integrated over crystal momentum for the field strength $F_0 = 0.07$ a.u. and $F_0 = 0.021$ a.u., showing the impact of field strength on the spectral features. The non-integer harmonic signals (blue arrows) are the result of strong-field-induced nonadiabatic coupling between the FBSs and electronic bands. Note that even-order harmonics can appear at individual
$k$ points (\textbf{b} and \textbf{c}), they cancel in the $k$-integrated
spectrum, which thus contains only odd harmonics. The observed non-integer peaks, therefore, arise from coupling involving FBSs. }
\label{fig1}
\end{figure*}

The emergence of the FBSB feature can 
be attributed to strong nonadiabatic dynamics enabled by the large Floquet parameter achieved in the experiment ($\mathcal{F}=4.2$). In this regime, nonadiabatic coupling and transient band renormalization between conduction bands and FBSs can introduce a dynamical energy gap, producing a weak transition channel where the TRS is transiently broken, analogous to Floquet engineering. 

%
\section*{TDSE evidence of Floquet-Bloch sidebands in HHG}
\begin{figure*}[ht!]
\centering\includegraphics[width=1\textwidth]{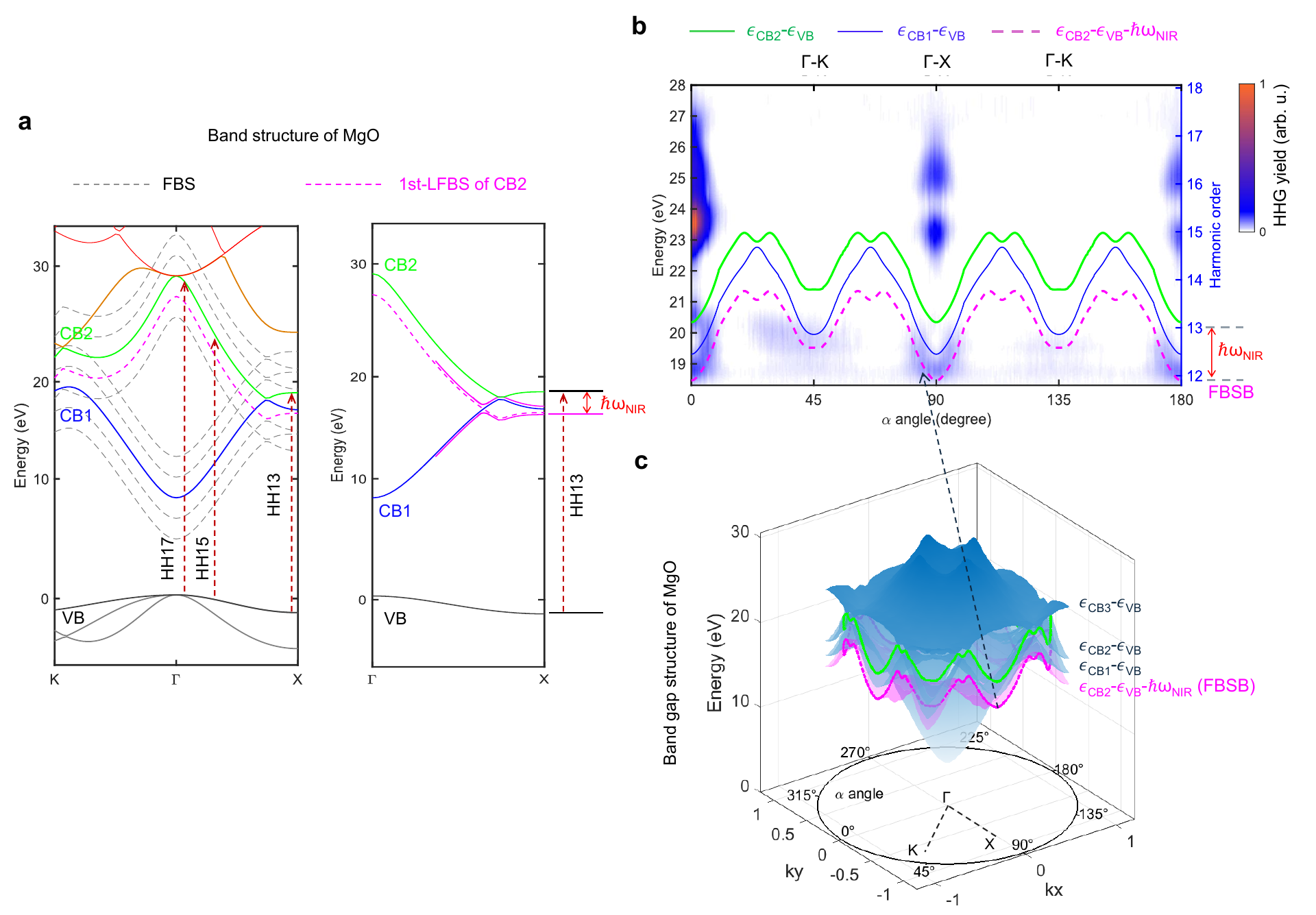}
\caption{\textbf{a}~Band structure of MgO along the $\Gamma$-X and $\Gamma$-K directions (left panel). The strong NIR driving field dresses the conduction bands with Floquet replicas (grey dashed curves), bringing the first-order lower FBS of CB2 (1st-LFBS, purple dashed curve) into near-degeneracy with CB1 (blue curve). The resulting avoided crossing enables strong nonadiabatic coupling between 1st-LFBS and CB1, generating coupled states (purple curves, right panel) and establishing a Floquet-assisted transition channel responsible for the observed HH12 feature (an FBSB). The orange arrow indicates the additional resonance condition for HH13, HH15 and HH17, where $\epsilon_{\mathrm{CB2}}-\epsilon_{\mathrm{VB}}$ becomes resonant with HH13 at the BZ edge. 
\textbf{b} Orientation dependence of MgO band gaps computed from DFT, including gap~1 ($\epsilon_{CB1}-\epsilon_{VB}$, blue curve), gap~2 ($\epsilon_{CB2}-\epsilon_{VB}$, green curve) and the energy gap involving the 1st-LFBS (FBSB, $\epsilon_{CB2}-\epsilon_{VB}-\hbar\omega$), purple dashed curve). The energy gap involving 1st-LFBS, around $\Gamma$-X and $\Gamma$-K directions, reproduces the arc-like morphology observed in the experimental $\alpha$ scan, demonstrating that the spectral feature of FBSB (located at HH12 in the $\Gamma$-X direction) inherits the momentum-dependent structure of the Floquet-dressed bandgap near the BZ edge. 
\textbf{c} Comparison of the arc-like feature with calculated 3D bandgap structures. The dashed purple curve shows the energy gap involving 1st-LFBS, reproducing the arc-like structure's shape.}
\label{fig31}
\end{figure*}

To understand the microscopic origin of the experimentally observed feature, we analyze the HHG response using TDSE simulations. The results of the TDSE simulations are shown in Fig.~\ref{fig1} for different values of the crystal momentum, $k$, ranging from $-\pi/a$ to $\pi/a$, where $a$ denotes the lattice constant (see Methods for details
).
First, the field-free band structure of our model potential is shown in Fig.~\ref{fig1}a, including the valence band (VB, black) and three conduction bands—CB1 (blue), CB2 (green), and CB3 (yellow). The FBSs of CB1 and CB2 are plotted with purple dashed curves as they contribute most to HHG. The calculated HHG spectra (laser frequency $\omega=0.057$ a.u., corresponding to 800 nm with 14 optical cycles and a 15~fs FWHM pulse envelope to match the experiment) as a function of crystal momentum~$k$ are shown in Fig.~\ref{fig1}b and Fig.~\ref{fig1}c, where the formation of FBSBs (purple dashed curves) is clearly captured. While the system response is integrated over momentum (see Methods section), it is informative to present the k-resolved harmonic spectra. Three interband transition paths are identified, corresponding to three band gaps: 
gap~1 ($\epsilon_{\mathrm{CB1}} - \epsilon_{\mathrm{VB}}$, blue), 
gap~2 ($\epsilon_{\mathrm{CB2}} - \epsilon_{\mathrm{VB}}$, green), 
and gap~3 ($\epsilon_{\mathrm{CB3}} - \epsilon_{\mathrm{VB}}$, yellow). 
The field-induced FBSBs trace the band-gap dispersions of gap~1 and gap~2, forming multiple photon-replica branches. Under this condition, the FBSBs remain clearly distinguishable with a bandwidth of roughly 1 eV, indicating that FBSs are formed, although the induced steady state is not yet reached.

Furthermore, in Fig.~\ref{fig1}a, near the BZ edge (in black dashed square), multiple crossings between the FBSs and conduction bands appear. These crossings suggest that nonadiabatic coupling or band-structure hybridization may take place in this region. The extent of this hybridization depends on the Floquet parameter~\cite{dunlap1986dynamic,ikeda2018floquet}, which characterizes the strength of the light–matter interaction:  
\begin{equation}
    \mathcal{F} = \frac{e F_0 a}{\hbar \omega}=\frac{\omega_B}{\omega},
\end{equation}
where $F_0$ is the peak electric field strength, $\omega$ is the laser angular frequency, $\omega_B$ is the Bloch frequency and $e$ is the elementary charge. The Floquet parameter also measures how rapidly electrons traverse the BZ compared with the optical cycle, establishing a direct connection between Bloch transport and nonadiabatic Floquet dynamics.

When $\mathcal{F}$ reaches 1, electrons can traverse the entire BZ, entering the nonadiabatic regime where FBSs start to form. As shown in Figure~\ref{fig1}b, we choose a peak field strength of $F_0 = 0.007$ a.u. (laser peak intensity 1.72$\times$10$^{12}$W/cm$^2$), corresponding to a Floquet parameter $\mathcal{F} =0.98$. The constructed FBSBs intersect with successive harmonic orders, revealing nonadiabatic coupling between the FBSs and the conduction bands~(CB1 and CB2), especially at the BZ edge. This behavior reflects the onset of light-dressed hybridization between different energy states. 
The resulting strong-field nonadiabatic coupling breaks TRS and gives rise to additional spectral features in the total HHG spectrum, beyond single harmonic orders, as illustrated in Fig.~\ref{fig1}d~(blue). 

When $\mathcal{F} > 1$, electrons can travel across multiple BZs \cite{catoire2015above}, and the nonadiabatic coupling between electronic states becomes increasingly significant. Figure~\ref{fig1}c shows the same spectrum with a larger peak field strength, $F_0 = 0.021$ a.u. (laser peak intensity 1.60$\times$10$^{13}$W/cm$^2$), corresponding to a Floquet parameter $\mathcal{F} = 2.94$. In this case, the nonadiabatic coupling between the FBSs and electronic bands becomes more pronounced. As more FBSBs are generated, their overlap with the three band gaps becomes intrinsic. The total HHG spectrum (Fig.~\ref{fig1}d, orange) exhibits many transition spectral peaks induced by this nonadiabatic process. This feature is more discernible between HH6 and HH15 (blue arrows), where interband polarization is mostly contributed by the BZ edge. 
Unlike the previous study~\cite{schmid2021tunable}, these signals, observed in the energy range of even and non-integer harmonics after integration over all crystal momenta, arise from strong-field-induced nonadiabatic coupling between the FBSs and the electronic bands, rather than from surface states, which are absent from the initial model.

In addition, as the field strength $F_0$ increases, the corresponding Keldysh parameter, $\gamma_K = \frac{\omega\sqrt{2\Delta\epsilon_g}}{F_0}$ ($\Delta\epsilon_{\text{g}}$ denotes the band gap: $\epsilon_{CB1}-\epsilon_{VB}\mid_{k=0}$), decreases. In our calculations, when the field strength $F_0 = 0.007$ a.u., $\gamma_K = 4.3$, indicating that multi-photon transitions dominate the HHG process and occur in the perturbative regime. When the field strength is increased to $F_0 = 0.021$ a.u., $\gamma_K  \approx 1.0$, suggesting that electron tunneling processes begin to occur, which is more favorable for the formation of FBSs~\cite{saathoff2008laser}. This condition closely matches the experiment, for which $\gamma_K  \approx 1.0$. Moreover, although FBS-induced sidebands are resolved in the calculation, their finite spectral bandwidth indicates that the FBSs are not fully established as stationary states, reflecting their transient nature.

\section*{Nonadiabatic coupling near the BZ edge}
%
\begin{figure*}[ht!]
\centering\includegraphics[width=1\textwidth]{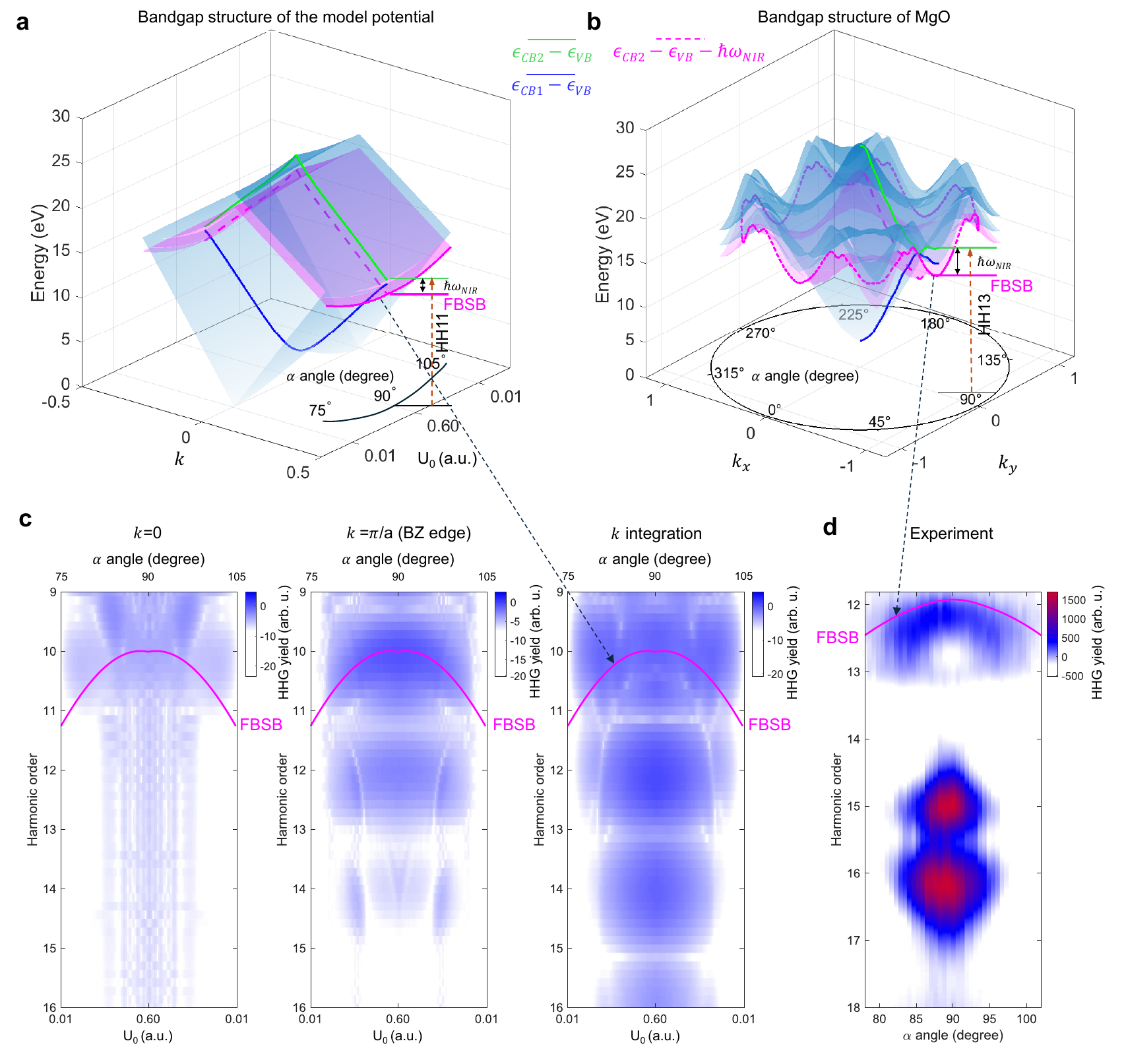}
\caption{Numerical simulation of nonadiabatic coupling and its experimental correspondence in MgO. 
\textbf{a}~Bandgap structure of the 1D model potential: gap~1 and gap~2 (blue surfaces), and the first-order FBSB of CB2 (purple surface). Cross-section at $U_0=0.60$~a.u. shows gap~1 (blue curve), gap~2 (green curve), and the FBSB of CB2 (purple dashed curve). Their crossing near the BZ edge indicates nonadiabatic coupling, giving rise to a resonant transition at approximately 16 eV. 
\textbf{b}~DFT bandgap structure of MgO showing a similar crossing between gap~1 and the first-order FBSB near the BZ edge as \textbf{a}, corresponding to a resonance at $\sim$19~eV.
\textbf{c}~Simulated HHG spectra from the 1D TDSE model. Left: fixed crystal momentum $k=0$ showing less contribution of FBSs. Middle: fixed crystal momentum $k=\pi/a$ showing multiple Floquet--Bloch replicas. Right: $k$-integrated spectrum where only the first-order FBSB survives, appearing one photon energy below HH11, confirming its nonadiabatic origin.
\textbf{d}~Experimental HHG spectra versus crystal azimuthal rotation $\alpha$. The first-order FBSB arc is imprinted around HH12, below HH13, consistent with nonadiabatic coupling to the conduction band and in agreement with the simulation.}
\label{fig4}
\end{figure*}

The combined experimental observations and TDSE results indicate that the feature---an FBSB near HH12---along the $\Gamma$--X direction originates from nonadiabatic coupling between CB1 and an FBS near the BZ edge. We now analyze the physical origin of this feature within a dressed-state picture based on the DFT band structure of MgO.

Under the high peak field of the NIR driver, the electronic wave packet is promoted into both the first and second conduction bands (obtained from DFT calculations; see Methods for details). Along the $\Gamma$–X direction, CB1 and CB2 approach each other near the BZ edge, forming an avoided-crossing-like configuration where their energy separation falls below a single NIR photon energy. The periodic driving field further dresses these bands with Floquet replicas (FBSs, grey dashed curves in Fig.~\ref{fig31}a), creating near-resonant conditions for Floquet-mediated interband coupling.

In this regime, the dominant transition channel involves the first-order lower FBS of CB2 (1st-LFBS, purple dashed curve), which lies one photon energy below CB2 and intersects CB1 near the BZ edge, as shown in the right panel of Fig.~\ref{fig31}a. At this quasi-degeneracy point, strong nonadiabatic coupling is triggered by the driving field, enabling nonadiabatic hybridization between the two states. The resulting dressed states open a dynamical energy gap separated by one laser photon energy, thereby facilitating a Floquet-assisted interband transition that provides a robust transition pathway. This mechanism manifests experimentally as the emergence of the FBSB located in the energy range corresponding to HH12 signal. The nonadiabatic coupling is described by the interaction Hamiltonian for two states
\begin{equation}
    H_\mathrm{couple}=
    \begin{pmatrix}
    \epsilon_\mathrm{CB2}-\hbar\omega_\mathrm{NIR} & V \\
    V^{\dagger} & \epsilon_\mathrm{CB1} \
    \end{pmatrix} \, ,
\label{eq:Ham}
\end{equation}
where $\epsilon_\mathrm{CB1}$ and $\epsilon_\mathrm{CB2}$ are the eigenenergies of CB1 and CB2, $\epsilon_\mathrm{CB2}-\hbar\omega_\mathrm{NIR}$ refers to the 1st-LFBS of CB2, and $V=\bra{\mathrm{CB1}}\mathbf{A} \cdot\mathbf{p}\ket{\mathrm{CB2}}$ with $\ket{\mathrm{CB1}}$ and $\ket{\mathrm{CB2}}$ denoting the two eigenstates. 
Diagonalising the Hamiltonian~\eqref{eq:Ham}, we obtain the coupled bands~(purple curves in the right panel of Fig.~\ref{fig31}a). These coupled states define new transition pathways in HHG, which can be directly mapped onto specific harmonic orders.

These two coupled states manifest in the HHG within the 18--21 eV window, corresponding to the transitions
$\epsilon_{\mathrm{CB2}}-\hbar\omega_{\mathrm{NIR}}-\epsilon_{\mathrm{VB}}$ and 
$\epsilon_{\mathrm{CB2}}-\epsilon_{\mathrm{VB}}$, respectively. 
At the BZ edge, HH13 becomes nearly resonant with the energy gap 
$\epsilon_{\mathrm{CB2}}-\epsilon_{\mathrm{VB}}$ (Fig.~\ref{fig31}a, orange dashed arrow), leading to a distinct emission in the vicinity of HH13. 
In contrast, higher order harmonics such as HH15 and HH17 are primarily dominated by interband polarization near the $\Gamma$ point, where the 1st-LFBS and CB1 crossing does not occur (as shown in the left panel of Fig.~\ref{fig31}a). 
Consequently, the nonadiabatic process can not be efficiently probed by HH15 and HH17, making the nonadiabatic crossing feature significantly weaker or absent in these harmonic orders. In addition, this feature is likely associated with a similar nonadiabatic coupling process arising from the interaction between CB2 and a FBS of CB3 near the $\Gamma$ point, where a transition peak appears at HH16. However, as the present work focuses on CB1–CB2 dynamics, the CB3-related channel is not discussed further here.

To verify our interpretation of the arc-like structure observed in the $\alpha$-angle scan of the HHG spectrum, we compare the measured spectrum with MgO's band gaps at the BZ edge calculated using DFT approach (Fig.~\ref{fig31}b,c).
One can notice that the first lower-order FBSB of the CB2, shown with a dashed purple curve, matches the shape of the arc-like structure. This matching suggests that the morphology of the second bandgap structure of MgO~($\epsilon_{CB2}-\epsilon_{VB}$) at the BZ edge is imprinted in the HHG spectrum $\alpha$ scan assisted by this unique nonadiabatic process, resulting in the arc-like structure. This confirms that HH12 along the $\mathrm{\Gamma}$-X direction originates from an FBSB mediated by nonadiabatic coupling. Despite not occurring at the central $\Gamma$ point, this process exhibits Floquet-engineered behavior similar to that previously observed in TrARPES measurements.

\section*{Reproducing the arc-like HHG feature}

While the dressed-state model captures the dominant coupling pathway, it does not account for the full temporal dynamics of the FBSB. We therefore performed complementary one-dimensional TDSE simulations in order to reproduce the arc-like feature observed in the experiment and further analyze it.

To emulate the variation of the $\alpha$ angle scan in the experiment, we parametrically tune the lattice potential depth $U_0$ from 0.01 to 0.60 a.u., as it effectively modifies the band structure in a manner analogous to rotating the crystal lattice, thereby mimicking the $\alpha$-dependent band dispersion that governs the HHG emission in the experiment~(see Methods section for details).

The bandgap structure of the model potential is shown in Fig.~\ref{fig4}a, as it directly determines the photon emission energies in HHG. The fundamental interband gaps, gap~1 ($\epsilon_{CB1}-\epsilon_{VB}$) and gap~2 ($\epsilon_{CB2}-\epsilon_{VB}$), are plotted as blue surfaces. The first-order FBSB of CB2 ($\epsilon_{CB2}-\epsilon_{VB}-\hbar \omega$) is shown as the purple surface. For clearer comparison, a cross-section at $U_0=0.60$ a.u. is presented, where gap~1 (blue curve), gap~2 (green curve), and the FBSB of CB2 (purple dashed curve) are overlaid. The intersection between gap~1 and the FBSB near the BZ edge indicates strong nonadiabatic coupling between the 1st-LFBS and CB1, resulting in a resonant transition at approximately 16~eV (near HH10).

The arc-like feature originates from the first-order FBSB of CB2 at the BZ edge (purple curve), which embodies the spectral imprint of nonadiabatic coupling and evolves with the lattice potential depth $U_0$. At the BZ edge, HH11 becomes the key harmonic channel probing this feature, as its photon energy is resonant with gap~2. The calculated HHG spectra are shown in Fig.~\ref{fig4}c. The FBSB signature appears both at fixed crystal momentum $k=\pi/a$ (BZ edge, middle) and after full $k$-integration (right), located one pump photon energy below HH11. By contrast, the contribution at the fixed crystal momentum $k=0$ ($\Gamma$ point, left) is much weaker, indicating that this nonadiabatic coupling is predominantly driven by dynamics near the BZ edge. At $k=\pi/a$, two Floquet-Bloch replicas are visible (around HH10 and HH12), whereas only the first-order FBSB persists after $k$-integration. This persistent contribution confirms that first-order FBSB can be observed experimentally.

The experimental observations are quantitatively reproduced by our simulations. Fig.~\ref{fig4}b presents the bandgap structure of MgO, where the crossing between gap~1 and the first-order FBSB near the BZ edge indicates strong nonadiabatic coupling between the 1st-LFBS and CB1, leading to a transition at approximately 19~eV. Under this condition, HH13 becomes the dominant harmonic channel probing this coupling effect. The corresponding first-order FBSB arc feature, located one photon energy below HH13, is imprinted in the HHG spectrum at HH12 and evolves with $\alpha$ angle, as shown in Fig.~\ref{fig4}d, with a precise $\alpha$ scanning by a step of 0.1$^{\circ}$. Although the specific harmonic orders differ between experiment and calculation because of differences between the MgO and model band structures, our model captures the same underlying physics, indicating that this behavior reflects a universal feature of the driven electron dynamics.

\section*{Conclusions}

We demonstrated clear evidence of Floquet engineering in the strong-field regime, captured in the HHG spectra emitted by a wide-bandgap insulator. In this regime, when the MgO crystal is driven by a strong few-cycle near-infrared field, we observed pronounced manifestations of Floquet–Bloch sidebands arising from the laser dressing of the conduction bands. Notably, these Floquet-engineered features emerge in the emitted HHG signal at harmonic orders whose photon energies coincide with the band gap near crossing points at the Brillouin zone (BZ) edge of the dressed band structure, establishing an indirect observation of the laser-dressed electronic structure in the emitted high-harmonic spectra. We clearly observed this behavior in experiments, where it appears as a distinct arc-like feature in the dependence of harmonic emission on the crystal orientation.
This behavior highlights the sensitivity of HHG to laser-dressed band morphology and nonadiabatic Floquet dynamics.
Our time-dependent Schrödinger equation simulations reproduce these observations and attribute the anomalous harmonic response to laser-induced band dressing and strong nonadiabatic coupling near the BZ edge. Together, experiment and theory identify Floquet-mediated coupling as the microscopic origin of the observed harmonic features and reveal a dynamical breaking of time-reversal symmetry under strong-field driving.
Our results establish Floquet engineering as a robust and powerful framework for probing and controlling extreme nonlinear light–matter interactions in solids. Beyond identifying field-dressed electronic dynamics, this approach opens a promising route toward attosecond-resolved structural and electronic tomography in condensed-matter systems, as well as toward dynamical band-structure engineering and field-controlled material properties.


\section*{Methods}

\subsubsection*{Material characteristics}
In the experiment, we utilize a $z$-cut MgO monocrystal~(100 orientation) 
of $\SI{1}{cm} \times \SI{1}{cm}$ size as the target for HHG. The surface of MgO is two-sided, optically polished, and the thickness is measured to be $\SI{100}{\mu m}$. MgO crystallises in the rock-salt~(NaCl) structure belonging to the cubic crystal system with the space group Fm$\overline{3}$m. Its unit cell is face-centered cubic~(FCC) with a lattice parameter of approximately \(a = 4.21 \, \text{\AA}\). The Mg--O bond length is about \(2.1 \, \text{\AA}\), and the structure is highly symmetric with a density of approximately $\SI{3.58}{g/cm^3}$ under standard conditions. The measured bandgap energy of MgO is about \SI{7.8}{eV}.

\subsubsection*{Experimental setup}

We measure HHG spectra generated in a $\SI{100}{\mu m}$ MgO monocrystal, cut along the~(100) plane. The experimental setup is shown in Fig.~\ref{fig2}a. The experiment was carried out at the ELI Beamlines Facility~\cite{hort2019high}, where a \SI{15}{fs}, sub-six-cycle pulse was focused onto the MgO crystal using a spherical mirror with a focal length of $f=\SI{5}{m}$. The HHG signal is filtered by a \SI{150}{nm} aluminium foil and characterised by a calibrated XUV flat-field spectrometer with a CCD detector.

A 3D motorised stage is used to control the azimuthal crystal angle $\alpha$ over $[0-180]^\circ$ including the $\mathrm{\Gamma}$-X~(at $90^\circ$) and $\mathrm{\Gamma}$-K~(at $45^\circ$) directions, with an increment of $0.5^\circ$. The HHG signal is detected in a transmission configuration.

The laser spectrum and temporal profile, as shown in the right panel of Fig.~\ref{fig2}a, delivers infrared \SI{15}{fs} pulses with \SI{16.8}{mJ} energy at \SI{1}{kHz} repetition rate by using a broadband optical parametric chirped-pulse amplification~(OPCPA) system. 4\% of the OPCPA energy is reflected by a wedge and used for HHG. The calculated laser peak intensity on target is $\SI{3e13}{W/cm^2}$, below the damage threshold of MgO that we measure to be $\SI{6e13}{W/cm^2}$, in agreement with~\cite{allegre2025extension}. 

In our experiment, interband HHG spectra are observed from the second plateau, in the energy window of 18-28 eV. In this energy range, HHG spectroscopy can probe the first two conduction bands (CB1 and CB2), as well as the emergence of their FBSs, and the nonadiabatic coupling between them.

\subsubsection*{DFT calculation of band structure}
\begin{figure}[h!]
\centering
\includegraphics[width=0.49\textwidth]{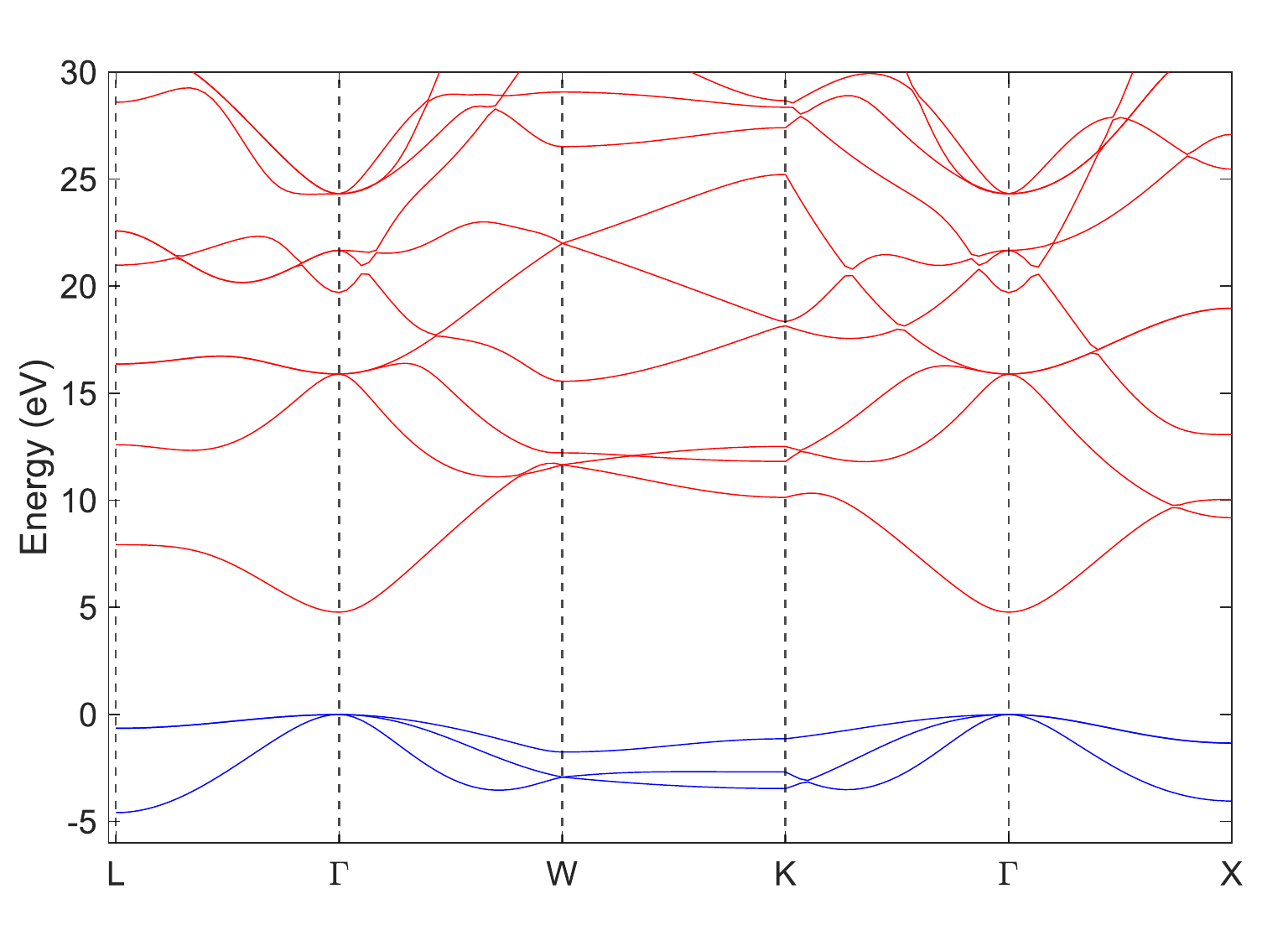}
\caption{MgO band structure calculated by DFT under LDA approximation. Red colour indicates CBs, while the blue colour refers to VBs.}
\label{fig:band}
\end{figure}

The band structure of the bulk MgO crystal is calculated using DFT~\cite{medvedev2017density} approach. The Octopus package~\cite{marques2003octopus, castro2006octopus} has been used for that purpose. The band structure is calculated using the adiabatic local-density approximation~(LDA)~\cite{kohn1965self} and employs SG15 norm-conserving pseudopotentials~\cite{Hamann2013, Schlipf2015, Scherpelz2016} (see Fig.~\ref{fig:band}).

It is important to note that within LDA, the band gap of semiconductors and insulators is underestimated and does not account for excitonic effects. We used the scissor operator for a better description of the band gap. Despite this discrepancy, the LDA approach correctly describes the dispersion of the VB and CBs, thus properly describing the dynamics of intraband current and interband polarisation.

\subsubsection*{TDSE calculation}
We numerically solve TDSE with a 1D model potential $V(x)$, which follows the Kronig-Penney model. In this model, the potential well is centred at $x = 0$ and has a width of $a/2$ with $a =7.8$ a.u. being the periodicity of the system in real space, mimicking MgO. The lattice potential value is set to be 1.10 a.u., which gives a band gap value of 7.8 eV.

The vector potential  $A(\omega,t)$ of the driving field has a $\cos^2$ envelope and is described as follows:
\begin{equation}
    A(\omega,t)=\frac{F_0}{\omega} \cos(\frac{1}{2N_c}\omega t)^2 \sin(\omega t)
\end{equation}
where $\omega$ is the laser angular frequency, $F_0$ is the field strength and $N_c$ denotes the laser cycle. In the experiment, the driving laser has a central wavelength of approximately 800 nm and a pulse duration of 15 fs (FWHM). To reproduce these experimental conditions in the simulation, we set the laser angular frequency to $\omega = 0.057$ a.u. and the pulse length to $N_c = 14$ optical cycles. The field strength $F_0$ is set to be 0.072 a.u., corresponding to a peak laser intensity of 3.2$\times10^{13}$W/cm$2$. The total pulse duration is 37 fs ($2\pi N_c/\omega$).

The wave function describing the system is given by the Bloch wave function $\Psi(x,t) = e^{ik_0x}u(k_0,x,t)$ with the reduced wave function u($k_0$,x,t), which is periodic in space with $2\pi/a$ periodicity, and the initial momentum $k_0$. The reduced part of the Bloch wave function is obtained by solving the TDSE in the dipole approximation in the velocity gauge. The TDSE is explicitly given by
\begin{equation}
    i\frac{\partial u}{\partial t} = \tilde H(k_0,t)u = \left\lbrace\frac{[\hat{P}+k_0+A(t)]^2}{2}+V(x)\right\rbrace u \, ,
\end{equation}
where $\hat{P}$ is the electron momentum operator. Using the velocity gauge ensures that we keep the periodicity of the Hamiltonian over time. 

The wave function is decomposed on a plane wave basis and is solved directly in real space. The propagation is performed using the Lanczos algorithm (with a basis of size $10$) with the initial condition $u(k_0,x,t = 0) = \varphi(k_0,x)$. $\varphi(k_0,x)$ is the eigenstate of the stationary Schr\"odinger equation with eigenvalue $\epsilon_\mathrm{n}(k_0)$, describing the band $n$. In this work, $\varphi(k_0,x)$ is the eigenstate for $n\geq 1$. 
Here we start with $n = 2$ to get the bandgap at the $\mathrm{\Gamma}$ point, same as for the DFT calculation. The numerical convergence is checked using the procedure of varying the grid parameters. Typically, 21 plane waves are used to fully converge the TDSE solution.

Once the wave function over time has been obtained, the HHG spectrum is calculated from the derivative of the current using the Ehrenfest theorem:
\begin{equation}
    \frac{\partial J(k_0,t)}{\partial t} =  \int dx \, u^\dagger (k_0,x,t) 
    \left[ \frac{\partial V}{\partial x} + E(t) \right] u(k_0,x,t) \, .
\end{equation}
The total current is then found by integrating over $k_0$ in the BZ, that is 
\begin{equation}
\frac{\partial J(t)}{\partial t} =\int_{BZ} dk_0 \, \frac{\partial J(k_0,t)}{\partial t} \, .
\end{equation}
Here, $k_0$ spans from $-\pi/a$ to $\pi/a$ using 201 grid points, corresponding to a step size of 0.01$\times\pi/a$, to reach the convergence of the total current.

The power spectrum of the emitted harmonics is then evaluated by taking the Fourier transform of the current as
\begin{equation}
    I_\mathrm{HHG}(\omega)=\left| \omega\int_{-\infty}^{\infty}dt \, e^{i\omega t}J(t) \right|^2 \, .
\end{equation}

As mentioned above, the calculations are performed in the velocity gauge to ensure momentum conservation. In Eq.~\eqref{eq:Ham} of the main body of the article, the dressed-state picture is also performed using the velocity gauge~(as the eigenvalues converge to the field-free states~\cite{vabek2022ionization}) as can be seen from the matrix elements $V$. Note that for the particular case of the first-order nonadiabatic coupling there is gauge invariance of the result.

\section*{Acknowledgments}
A.Z. acknowledges funding from UK Research and Innovation (UKRI)  under EPSRC-DTP and STFC-XFELHub, the Royal Society under the projects 'AMOS’~RGS\textbackslash{}R1\textbackslash{}211053 and ‘ASTRO’~IE2GR31\textbackslash{}3330. 
A.Z. and Z. D. acknowledge funding by the UK Research and Innovation (UKRI) under the UK government’s Horizon Europe funding guarantee [Grant No. EP/Z000807/1].
Portions of this research were carried out at the ELI Beamlines Facility, a European user facility operated by the Extreme Light Infrastructure ERIC. The experiments were conducted at the HHG Beamline and we acknowledge the L1 \& F-SYNC laser team.
M.K. acknowledges The Royal Society funding under URF\textbackslash R1\textbackslash 231460. 
Computer time for this investigation was provided by the computing facilities of the MCIA~(Mésocentre de Calcul Intensif Aquitain) and  the Computational Research, Engineering and Technology Environment~(CREATE) from King's College London.

\section*{Author contributions}
A.Z. supervised and funded the project; O.H., L.J., O.F., M.A. contributed to the development and operation of the HHG beamline; L.J. designed the measurement setup for the MgO, including sample holders and motorization stages; C.Z., L.J., X.Z. performed the experiment under the supervision of A.Z. and O.H.; F.C. developed the algorithm and implemented the calculation code; C.Z. used F.C.'s code to simulate the experiment under the supervision of F.C. and A.Z.; C.Z. and L.J. did the calibration of the experiment; C.Z., X.Z. analysed the data and prepared the figures; C.Z., A.Z., F.C., B.T.Q.M., M.K. wrote the manuscript with input from all authors. All authors participated in the discussions and interpretation of the data.

\section{Competing interests}

The authors declare no competing interests.

\balance


\bibliographystyle{arthur} 
\bibliography{sn-bibliography}{}

\hfill 

\end{document}